# Hall-effect and resistivity measurements in CdTe and ZnTe at high pressure: Electronic structure of impurities in the zincblende phase and the semi-metallic or metallic character of the high-pressure phases

D. Errandonea[1,3], A. Segura[1,2], D. Martínez-García[1,2], and V. Muñoz[2]

[1] Malta Consolider Team, [2] Departamento de Física Aplicada-ICMUV, [3] Fundación General de la Universidad de Valencia

Edificio de Investigación, C/Dr. Moliner 50, 46100 Burjassot (Valencia), Spain

**Abstract:** We carried out high-pressure resistivity and Hall-effect measurements in single crystals of CdTe and ZnTe up to 12 GPa. Slight changes of transport parameters in the zincblende phase of CdTe are consitent with the shallow character of donor impurities. Drastic changes in all the transport parameters of CdTe were found around 4 GPa, i.e. close to the onset of the cinnabar to rock-salt transition. In particular, the carrier concentration increases by more than five orders of magnitude. Additionally, an abrupt decrease of the resistivity was detected around 10 GPa. These results are discussed in comparison with optical, thermoelectric, and x-ray diffraction experiments. The metallic character of the *Cmcm* phase of CdTe is confirmed and a semi-metallic character is determined for the rock-salt phase. In zincblende ZnTe, the increase of the hole concentration by more than two orders of magnitude is proposed to be due to a deep-to-shallow transformation of the acceptor levels. Between 9 and 11 GPa, transport parameters are consistent with the semiconducting character of cinnabar ZnTe. A two orders of magnitude decrease of the resistivity and a carrier-type inversion occurs at 11 GPa, in agreement with the onset of the transition to the *Cmcm* phase of ZnTe. A metallic character for this phase is deduced.

## I. Introduction

Even if transport measurements have been widely used in high-pressure research and have been applied to semiconductor physics since the foundational period [1, 2], after the appearance of the diamond-anvil cell (DAC), optical measurements have certainly been the technique of choice in high-pressure semiconductor physics. However, in the last decade high-pressure transport experiments have shown to be a powerful tool to investigate pressure-driven changes in the electronic structure of semiconductors [3 – 5]. Cadmium telluride (CdTe) and zinc telluride (ZnTe) are well known II-VI semiconductors with a wide spectrum of technological applications; e.g. solar cells and x- and $\gamma$-ray detectors [6]. Both tellurides upon compression undergoes a series of structural phase transitions which have been well document in the literature [7 – 10]. Despite the growing interest in the physical properties of CdTe and ZnTe, very little information currently exists on how their electrical properties are affected by compression. Two-points resistivity measurements have been done in CdTe by Samara [11] and Minomura [12]. Four-points resistivity measurements have been also performed but only in powder-samples of CdTe using a DAC [13]. In the case of ZnTe, only two-points resistivity measurements have been carried out [14, 15]. All these studies suffer from the drawback of being strongly affected either by contact resistance problems or carrier scattering at grain boundaries on powder samples. Additionally, in some cases, contradictory results have been published [11, 13]. More important, fundamental transport parameters like the carrier concentration and mobility can only be obtained upon the performance of Hall-effect studies. In addition, important issues like the semiconducting or metallic character of the high-pressure phases remains still open. As a consequence of these facts, in order to provide accurate information of the pressure effects on the transport properties of CdTe and ZnTe we have performed resistivity and Hall-effect measurements in single-crystalline samples upon compression using a four-points technique. The experiments were performed in a quasi-hydrostatic set-up up to 12 GPa. In section II we will describe the experimental methods and the obtained results will be reported and discussed in section III.

## II. Experimental details

Single crystals here used were grown by the Bridgman method (CdTe) [16] and the cold travelling heather method (ZnTe) [17]. Samples for transport measurements were cut from ingots with a wire saw, ground with abrasive powder, polished with diamond paste, and finally chemically polished with a 5% Br : methanol solution and washed in distilled water. CdTe crystals were n-type with typical resistivity 1.5 $\Omega$ cm, electron concentration 7.6 $10^{15}$ cm$^{-3}$, and mobility 550 cm$^2$/V s at room temperature (RT). ZnTe crystals were p-type with typical resistivity 2.9 $10^4$ $\Omega$ cm, hole concentration 1.7 $10^{13}$ cm$^{-3}$, and mobility 12 cm$^2$/V s at room temperature.

Hall-effect and resistivity measurements under pressure up to 12 GPa were made with steel-belted Bridgman tungsten carbide (WC) anvils with a tip of 15 mm in diameter. The samples were contained using two annealed pyrophyllite gaskets (0.5-mm thick each) in a split gasket geometry. The internal diameter of the gaskets was 5 mm and hexagonal boron nitride (BN) was used as pressure-transmitting medium. In our device, a 150 ton oil press is employed to apply the load in two opposing WC anvils. Samples were typically 200-µm thick and 3 × 3 mm$^2$ in size (sample 1). The pressure applied to the sample was determined by the calibration of the load applied to the anvils against high-pressure resistivity transitions in calibrants [18]. The maximum pressure achieved with this set-up is 13 GPa [19]. Experiments were repeated two times for each material to check their reproducibility. A schematic view on the experimental set-up and a picture of a loaded sample are shown in Fig. 1. Additional details of this set-up can be found in Ref. [18]. In CdTe, additional measurements constrained to 4 GPa were performed using larger WC anvils (27-mm diameter tip). In this case, the sample size was 5 × 5 mm$^2$ (sample 2) and the pressure medium was sodium chloride. More details about this second high-pressure device can be found in Ref. [4]. In CdTe the experiments were performed for pressure increase and release. In ZnTe we only report results for pressure increase since in the two experiments performed on it we suffered a contact breakage at 12 GPa.

In order to perform the transport measurements, four indium contacts were vacuum evaporated on the corners of the samples in the van der Paw configuration taking care that the contact size was always much smaller than the distance between the contacts (see Fig. 1). Silver (Ag) wires of 100 μm in diameter were used as electrical leads. To guarantee the good quality of contacts, the silver wires were flattened and sharpened at the tip, being soldered with high-purity indium to the evaporated indium electrodes. To avoid inaccuracy problems in the determination of the Hall coefficient ($R_H$) coming from offset voltages, we acquired two sets of Hall measurements, one for positive and one for negative magnetic field (0.6 T) directions. The linearity of the ohmic voltages on the injected current was checked out at different pressures. From the performed experiments we obtained the pressure evolution of the resistivity ($\rho$), the electron ($n$) or hole ($p$) concentration, and the carrier mobility ($\mu$).

### III. Results and discussion

### A. Cadmium Telluride

Fig. 2 shows the pressure dependence of $\rho$, $n$, and $\mu$ in a CdTe sample as obtained from our measurements up to 12 GPa (sample 1). Up to 3 GPa, the only noticeable change is a slight decrease of the electron mobility that, given the constancy of the electron concentration, results in a slight increase of the resistivity. The decrease of the electron mobility is likely to be mainly determined by the increase of the electron effective mass, correlated to the increase of the bandgap, according to the **k·p** model [20]. The pressure insensitivity of the electron concentration is consistent with the extrinsic conduction regime. All donor impurities are ionised at RT and the expected slight increase of their ionisation energy under pressure does not change the donor level occupation probability.

The transition to the rock-salt phase [7] is observed as abrupt changes in $\rho$, $n$, and $\mu$ at about 4 GPa, in agreement with x-ray diffraction experiments [7, 8]. Similar changes were observed in the experiments performed with the large anvils (sample 2), but in this case the onset of the changes took place at 3 GPa as can be seen in Fig. 2. This difference can be due, on the one hand, to the larger anvil surface, that allows for a more progressive change of the pressure and a more detailed exploration of the of the pressure range through which zincblende and cinnabar phases coexist. It is remarkable that changes in the transport parameters in this pressure range are fully reversible. On the other hand the much pour quasi-hydrostatic conditions of the second set of experiments can also play a role. This observation is consistent with the fact that the zincblende – cinnabar – rock-salt structural sequence is sluggish [7, 8, 21], being affected by the experimental conditions. Indeed, it is well-known fact, that changes of the pressure-transmitting medium can reduced by several GPa the onset of a given phase transition [22, 23]. It is worth noticing here that, at a smaller scale, in sample 1 the onset of the transition is announced by a progressive increase of the carrier concentration above 3 GPa. Subsequently, after the abrupt changes observed between 4 GPa and 4.5 GPa, a much slower increase is observed in $n$ up to 6 GPa. On the other hand, the mobility decreases from 4 GPa to 6 GPa, reaching a value of 3 cm$^2$/Vs. However, the increase of the carrier concentration is the dominant effect producing the decrease of $\rho$ reported in Fig. 2. Beyond 6 GPa the three transport parameters, $\rho$, $n$, and $\mu$, remain nearly constant up to 10 GPa. Above this pressure, the Hall effect could not be measured because the Hall voltage became very weak. Nevertheless, an additional decrease of the resistivity, of more than one order of magnitude, could be detected in CdTe at 10 GPa, in agreement with the occurrence of the rock-salt to *Cmcm* phase transition detected in x-ray diffraction experiments [7]. All the changes found in the transport parameters are reversible. However a large hysteresis is observed for the changes induced at 4 GPa. This is not strange since on pressure release the zincblende phase of CdTe is only recovered at 2 GPa according with x-ray diffraction experiments [8]. It is important to note here that from the decrease of $\rho$ found at 10 GPa, an electron concentration of the order of $5 \times 10^{22}$ cm$^{-3}$ is estimated for the *Cmcm* phase of CdTe, confirming than in this phase CdTe has a metallic character. This results is consistent with low thermoelectric power values ($S < 30$ µV/K) measured above 10 GPa for the *Cmcm* phase [24].

We would like to compare now our results with previous electrical measurements. Samara *et al.* [10] and Shchennikov *et al.* [24], using a two-points technique and single crystals of CdTe, observed a qualitatively similar pressure behaviour for $\rho$ than us, finding two sharp changes in the resistivity at 3.8 GPa and 10 GPa. However, He *et al.* [13], from their powder-samples experiments, reported a quite different evolution of $\rho$ upon compression. These authors measured a continuous decrease of $\rho$ with pressure, with three small inflexions at 4 GPa, 6.5 GPa, and 10 GPa. We think that the resistivity decrease reported by these authors below 4 GPa is an experimental artefact which can be quite possible related to contact problems in their powder samples, which is confirmed by the fact that they do not see any abrupt change in $\rho$ at the rock-salt to *Cmcm* transition (10 GPa). Indeed, the additional changes these authors reported for the resistivity at 15 GPa, 22.2 GPa, and 30 GPa [13] are smaller than the scattering of their data and similar to the experimental uncertainties. Therefore their attribution to electronic transitions is not experimentally supported.

In order to provide a deeper interpretation of our results we will compare them with optical-absorption measurements [25, 26]. In these experiments, the transition to the rock-salt phase is observed at 3.8 GPa as a drastic decrease of the sample transmittance. In particular, it was found that the sample becomes virtually opaque between 3.9 GPa and 4.5 GPa. At this pressure, a range of relative transparency is observed between 1.2 eV and 2.2 eV. As pressure increases up to 10 GPa, the sample transmittance increases in the whole transparency range and the overall transmitted intensity increases by more than one order of magnitude. Different behaviours were observed in the low- and high-energy tails of the transparency range. While the low-energy tail tends to saturate and remains virtually constant above 6 GPa, the high-energy edge moves monotonously to higher photon energies as pressure increases [25]. Above 10 GPa the transparency region gradually shrinks and disappears at about 11 GPa [25]. These results were explained on the basis of the density-functional theory (DFT) band-structure calculations performed by Güder *et al.* assuming that rock-salt CdTe is an indirect semiconductor with a small band-gap [25]. DFT calculations predicts also a similar behaviour for rock-salt ZnS and ZnSe [27, 28]. However, on the basis of our Hall-effect measurements and previous reflectance studies [25], the low-gap model would be confirmed only under the assumption that the electrons filling the conduction-band minimum states in rock-salt CdTe are generated by extrinsic donor defects created at the phase transition. Given the constancy of the electron concentration above 6 GPa (around $10^{21}$ cm$^{-3}$), as measured in two different samples, as well as the reversibility of the electron concentration on decompression, it seems more reasonable to assume that the free carriers have an intrinsic origin. This assumption necessarily leads to a semi-metallic character for the rock-salt phase of CdTe. Such high-electron concentrations would not be observed in an intrinsic semiconductor at room temperature, even with a band-gap energy as low as a few tens of meV. Fig. 3 shows a sketch of the proposed band structure of rock-salt CdTe around the Fermi level, based on the band structure calculations reported in Ref. [25]. Given the symmetry of the different points of the face-centred-cubic Brillouin zone, the conduction band consists of three equivalent valleys at the point X, while the valence band consists of four equivalent valleys at point L and twelve additional valleys in the GK direction. The density of states in the valence band is expected to be much larger than the one in the conduction band. This is the reason why the Fermi level is assumed to be at a larger energy from the bottom of the conduction-band minimum in the band sketch of Fig. 3.

The previous discussion also rules out the non-linear behaviour deduced for the band-gap ($Eg$) of rock-salt CdTe from the pressure dependence of $\rho$ in Ref. [13] assuming $\rho \propto e^{Eg/2kT}$, where k is the Boltzmann constant and $T$ the temperature. Such a direct correlation between resistivity and $E_g$ can be expected only in an intrinsic semiconductor. In zincblende CdTe ($Eg$~1.4 eV) the intrinsic carrier concentration is so low at RT that intrinsic samples would be virtually insulating. Free electrons are of extrinsic origin. Besides that, inter-grain barriers in pellet samples preclude any attempt of relating resistivity changes to intrinsic parameters.

**B. Zinc Telluride**

Fig. 4 shows the pressure dependence of $\rho$, $p$, and $\mu$ in ZnTe sample as obtained from our measurements up to 12 GPa. There it can be seen that up to 5 GPa $\rho$ smoothly decreases with pressure, mainly due to the increase of the hole concentration that compensates the slight monotonous decrease of the hole mobility. From 5 GPa to 7 GPa an one order of magnitude decrease of the resistivity is observed, due again to an increase of the hole concentration by two orders of magnitude, partially compensated by a decrease of the hole mobility by a factor 10. From 7 GPa to 9 GPa both the hole concentration and mobility remain constant. At 9 GPa a rise of $\rho$ starts, corresponding to a decrease of both the hole concentration and mobility, reaching the resistivity a maximum around 10 GPa. Beyond 11 GPa resistivity undergoes a very sharp decrease, associated to carrier-type inversion and a quick increase of the electron concentration. These results are qualitatively in agreement with those reported by Ohtani *et al.* [14] and Ovsyannikov *et al.* [15]. The carrier-type inversion was not detected in previous studies because the resistivity is insensitive to the sign change of the dominant charge carrier.

The values of the transport parameters at ambient pressure reveal the compensated character of these samples, consistently with photoluminescence (PL) measurements [29], in which PL peaks between 540 and 600 nm were attributed to a large concentration of donor acceptor pairs. The value of the hole mobility (about 12 cm$^2$/Vs) is much lower than the intrinsic one (about 60 cm$^2$/Vs) [30], which indicates ionised impurity concentrations between $10^{18}$ and $10^{19}$ cm$^{-1}$. As the hole concentration is about 2 $10^{13}$ cm$^{-3}$, the ionised donor and acceptor concentrations must be of the same order $N_A \geq N_D \approx 5\ 10^{18} cm^{-3}$. In p-type compensated semiconductors, from the charge neutrality equation, it can be deduced that the hole concentration is given by the following expression:

$$p = \frac{N_V(N_A - N_D)}{2N_D} e^{-\frac{\Delta E_A}{kT}} \qquad (1)$$

where $N_V$ is the effective density of states in the valence band and $\Delta E_A$ is the acceptor ionization energy. PL results indicate the presence of acceptor levels with ionization energies up to 200 meV [29].

Let us discuss how changes in the transport properties induced by pressure can be correlated to changes in the electronic structure.

i) As discussed above, the hole mobility is determined by ionised impurity scattering, and its decrease under pressure can be qualitatively explained. According to the **k·p** model [20] the effective masses of both heavy and light holes increase as the bandgap increases [31 - 33]. Also, the static dielectric constant of ZnTe decreases under pressure [33]. For ionised impurity scattering, both effects contribute to the decrease of the hole mobility under pressure.

ii) The ionisation energy of shallow acceptors increases under pressure, due to the increase of the light hole effective mass and the decrease of the static dielectric constant [33]. This would necessarily lead to a decrease of the hole concentration under pressure, opposite to what is actually observed. Consequently, free holes must be generated by the ionisation of deep acceptors and the increase of the hole concentration must be associated to a decrease of their ionization energy under pressure. Between ambient pressure and 5 GPa the hole concentration increases by a factor 10. Using Eq. (1), it is straightforward to determine the pressure coefficient of the acceptor ionization energy: $d\Delta E_A/dP = -11\pm2$ meV/GPa. This effect has been previously observed by other authors [34- 36] through PL measurements under high pressure. The deep character of ionised acceptors in Zn chalcogenides is attributed by these authors to a $C_{3v}$ distortion of the substitutional acceptors [34, 35] or the Zn vacancy [36]. Under pressure the distortion decreases and eventually the distorted configuration becomes unstable with respect to the non-distorted configuration with local $T_d$ symmetry. Remarkably, the pressure coefficient of the acceptor ionisation energy, as measured through PL, is of the order of -10 meV/GPa, very close to the one obtained in this work from transport measurements.

iii) At 6 GPa a further increase of the hole concentration is observed leading to a constant value of 6 $10^{15}$ cm$^{-3}$ up to 9 GPa. In the framework of the distorted local configuration, 6 GPa would be the pressure at which the distorted configuration becomes unstable and the deep acceptor levels become shallow. Let us check the consistency of this model by applying Eq. (1) to estimate the concentration of ionized donors $N_D$. The constant hole concentration between 6 and 9 GPa can be assumed to be $p = N_A-N_D$. The ionization energy at 6 GPa would be the ionization energy of shallow acceptors, about 25 meV, as calculated from the light hole effective mass and static dielectric constant at this pressure [33]. From the total change of the hole concentration we estimate that $\Delta E_A$ decreases by 145 meV between ambient pressure and 6 GPa. Then the deep-level ionization energy at ambient pressure would be about 170 meV. If we calculate the effective denstity of states at the valence band maximum from the heavy and light hole effective mass [33], we get an estimation of $N_D \sim 8$ $10^{18}$ cm$^{-3}$, close to the one obtained above from the hole mobility at ambient pressure. Let us finally comment on the strong decrease of the hole mobility, associated to the acceptor deep-to-shallow transformation. As the total ionized impurity concentration does not change, the mobility decrease should be associated to an increase of the scattering cross section. This is what is expected in a deep-to-shallow transformation, in which the localized potential of a deep level changes to a more extended Coulombian potential .

iv) Regarding the change in the transport parameters observed around 9 GPa, it can be related with the occurrence of a pressure-induced phase transition to the cinnabar phase detected by x-ray diffraction measurements [14, 15, 37, 38]. Cinnabar ZnTe is a semiconductor with a bandgap of some 1.2 eV [39]. The phase transition from zincblende to cinnabar is a first order one and the sample becomes polycrystalline. Carrier trapping and scattering at the grain frontiers explain the decrease of both the hole concentration and mobility, whose values around 10 GPa for ZnTe are close with those found for $n$ and $\mu$ in the the cinnabar phase of CdTe.

v) Regarding the decrease of the resistivity and the increase of the carrier concentration observed beyond 11 GPa, these changes can be related with the presence of precursor defects of the transition to the *Cmcm* phase of ZnTe [38]. It is reasonable to assume that the introduction of the precursor defects leads to the creation of a large concentration of donor centers, which first compensate the acceptors of p-type ZnTe and subsequently produce the observed carrier-type inversion, becoming the electron concentration:

$$n = \frac{N_C (N_D - N_A)}{2 N_A} e^{-\frac{\Delta E_D}{kT}} \quad (2)$$

where $N_C$ is the effective density of states in the conduction band and $\Delta E_D$ is the donor ionization energy. Note that this picture is also in agreement with the decrease of an order of magnitude found at 11 GPa in the carrier mobility. A value close to 0.2 cm$^2$/Vs is consistent with a highly-defective crystal. A low mobility was also found beyond 9 GPa for the cinnabar of CdTe indicating that this phase have a large concentration of defects too. It is worth commenting here that around 11 GPa a situation were both type of charge carriers, holes and electrons, contribute to transport properties could occur in a small pressure range. Under such situation the carrier concentration measured in Hall measurements ($n_H$) is:

$$n_H = \frac{(p\mu_h + n\mu_e)^2}{p\mu_h^2 - n\mu_e^2} \qquad (3),$$

where $\mu_h$ and $\mu_e$ are the mobilities of holes and electrons. Therefore, a precise determination of the pressure where the type-inversion takes place can be only made performing transport measurements at high-pressure over a temperature range.

vi) Finally, we would like to mention that the evolution of the electron concentration beyond 11 GPa suggests that as in CdTe, the *Cmcm* phase of ZnTe has a metallic character. If the values we obtained for *n* from 11 to 12 GPa are extrapolated to 15 GPa, a pressure at which the cinnabar-*Cmcm* transition is completed, a carrier concentration of around $10^{22}$ cm$^{-3}$ is obtained. This value is in agreement with that estimated from optical reflectance measurements [40]. The conclusion on the metallic character of the *Cmcm* phase is also in agreement with thermoelectric measurements previously performed in ZnTe [15, 23]. However, these studies were not conclusive regarding whether *Cmcm* ZnTe has a metallic hole- or electron-like conductivity. Our measurements support the second option.

**IV. Conclusions**

We reported high-pressure resistivity and Hall-effect measurements in n-type CdTe and p-type ZnTe up to 12 GPa. Continuous and reversible changes in the zincblende phase are consistent with the shallow character of donor levels in CdTe and a deep-to-shallow transformation of the acceptor levels in ZnTe. Concerning high-pressure phases, for CdTe, transport measurements have shown to be complementary with optical, thermoelectric, and x-ray measurements. In particular, the information provided by Hall-effect measurements has been crucial to decide between the low-gap semiconductor and semi-metal models for rock-salt CdTe, in favour of the second option. In the case of ZnTe, the transition to the cinnabar structure has been detected as abrupt changes in all the transport parameters. In addition, a carrier-type inversion and a sharp decrease of the resistivity were found at 11 GPa, being correlated with the onset of the transition to the *Cmcm* phase. For both CdTe and ZnTe it has been shown that the *Cmcm* phase has a metallic character with electron-like conductivity.

## Acknowledgments

This work was made possible through financial support of the MICINN of Spain under Grants No. CSD2007-00045, MAT2007-65990-C03-01, and MAT2007-66129 and by the Generalitat Valenciana under Grant No. GVPRE-2008-112.

**Figure Captions**

**Figure 1:** Schematic view of the opposed-anvils set-up used in the transport measurements. A picture of a sample loaded in a pyrophilite gasket, with contacts made on the Van der Pauw configuration, is shown together with a sample and pyrophilite gasket mounted in a steel-belted WC anvil.

**Figure 2:** Resistivity, carrier concentration, and mobility of CdTe at room temperature as a function of pressure. Solid symbols: upstroke. Empty symbols: downstroke. Squares: sample 1. Circles: sample 2.

**Figure 3 (colour online):** Proposed band-structure of semi-metallic rock-salt CdTe.

**Figure 4:** Resistivity, carrier concentration, and mobility of ZnTe at room temperature as a function of pressure. Different symbols correspond to different samples. The vertical dashed line indicates the carrier-type inversion.

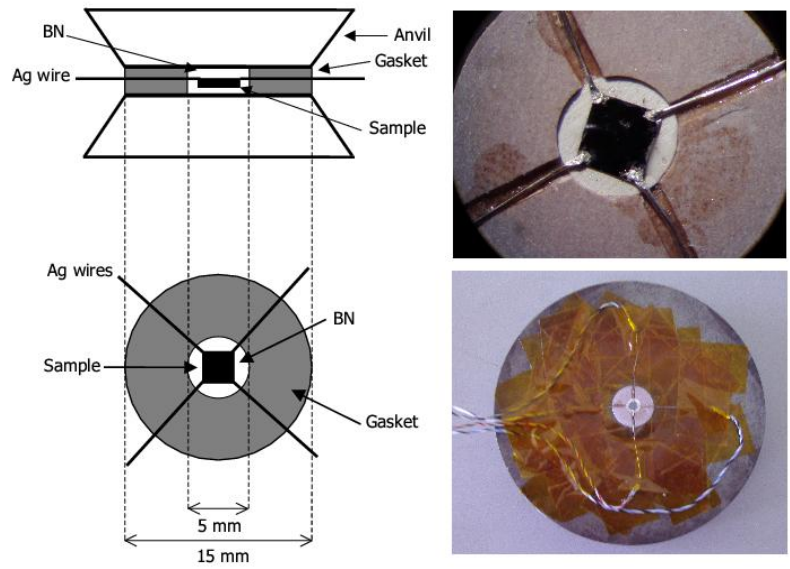

**Errandonea et al.**
**Figure 1**

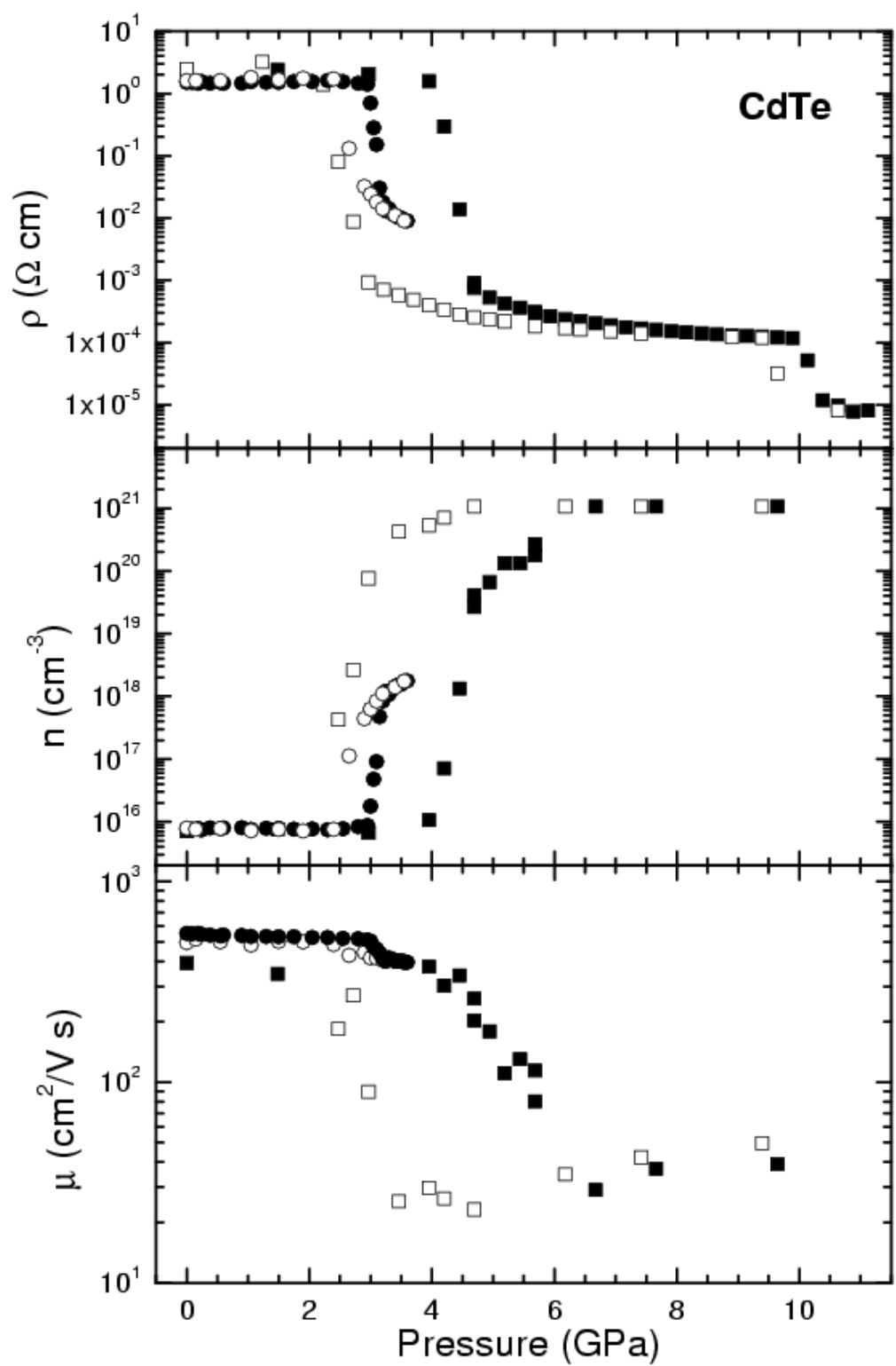

**Errandonea et al.**

**Figure 2**

 
18

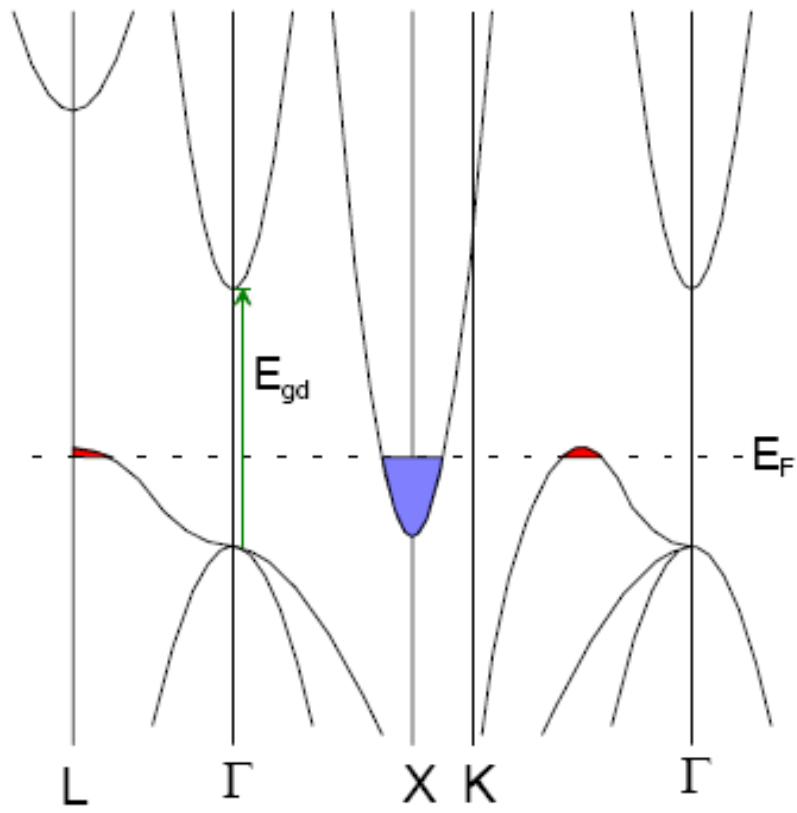

**Errandonea et al.**
**Figure 3**

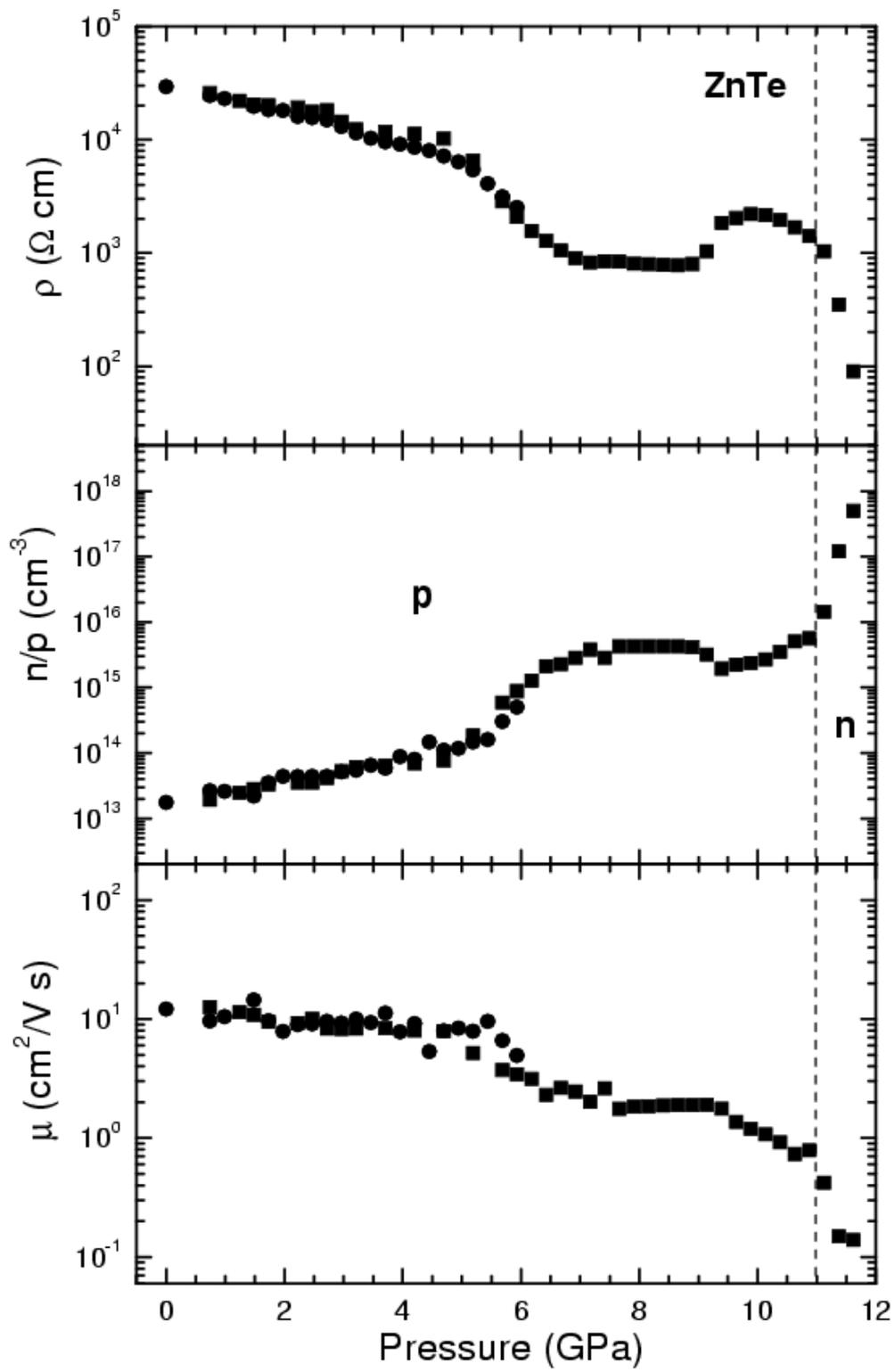

Errandonea et al.

Figure 4